\documentclass[twocolumn,amsmath,amssymb]{revtex4-1}
\usepackage{amsmath,amsfonts,amssymb,bm,hyperref,xcolor,amsopn}
\usepackage[capitalize]{cleveref}
\usepackage{graphicx,subfigure,epstopdf,multirow,diagbox}
\usepackage[normalem]{ulem}
\usepackage{soul}
\usepackage{cancel}
\soulregister\cite7
\soulregister\ref7

\begin{document}
\title{Solution landscape of droplet on rough surfaces: wetting transition and directional transport}
\author{Yuze Zhang$^{1}$}

\author{Xuelong Gu$^{1}$}

\author{Yushun Wang$^{1}$}

\author{Xianmin Xu$^{2}$}
\email{xmxu@lsec.cc.ac.cn}

\author{Lei Zhang$^{3}$}
\email{zhangl@math.pku.edu.cn}

\affiliation{$^{1}$School of Mathematical Sciences, Nanjing Normal University, Nanjing, 210024, China.\\$^{2}$LSEC, ICMSEC, NCMIS, Academy of Mathematics and Systems Science, Chinese Academy of Sciences, Beijing, 100190, China.\\$^3$Beijing International Center for Mathematical Research, Center for Quantitative Biology, Center for Machine Learning Research, Peking University, Beijing 100871, China.}

\begin{abstract}
Droplets on rough surfaces can exhibit various stationary states that are crucial for designing hydrophobic materials and enabling directional liquid transport. Here we introduce a phase-field saddle dynamics method to construct the solution landscape of wetting transition and directional transport on pillared substrates. By applying this method, we reveal the full range of Cassie-Baxter and Wenzel states, along with the complete wetting transition paths. We further elucidate the mechanisms of directional droplet transport on both hydrophobic and hydrophilic surfaces, demonstrating how surface design can influence directional movement.

\end{abstract}

\maketitle


Droplets on rough surfaces exhibit a range of intriguing phenomena \cite{Gennes1985,Gennes04,Bonn2009}, including the lotus effect \cite{Barthlott97,Koch09} and the rose petal effect \cite{Feng08}. In particular, various directional liquid transport behaviors have been observed recently on textured surfaces such as spider silk \cite{Zheng2010}, butterfly wings\cite{Zheng2007}, Sarracenia trichomes \cite{Chen2018}, and Araucaria leaves \cite{wang2021}. These phenomena are essential for numerous practical applications, including self-cleaning materials \cite{Quere08,Tuteja07,Wang20}, anti-icing surfaces \cite{Kreder16}, and water harvesting devices \cite{Zheng2010,Park16}. The wetting properties of droplet on rough surfaces are typically described by the Cassie-Baxter (CB) equation \cite{CB} or the Wenzel equation \cite{W}, which correspond to the two most energetically stable states in different scenarios—one with air trapped underneath the droplet and the other without. However, in reality, droplet on rough surfaces exhibit a multitude of metastable states that cannot be fully captured by these two equations \cite{Wang2023}. The intricate transition paths between these states pose significant challenges for understanding the CB-Wenzel wetting transitions \cite{Sbragaglia07,Giacomello2012} and the associated directional liquid transport phenomena.

To understand wetting transition on rough surfaces, there are two theoretical approaches: a force-based approach \cite{Yoshimitsu02,Reyssat06,Quere08} and an energy-based approach \cite{Patankar04,Ishino04,Whyman12,Bormashenko15}. In the latter, a crucial aspect is estimating the energy barrier between the Wenzel and CB states. Quantitatively determining this energy barrier is challenging, as it requires identifying the transition path and the lowest energy barrier connecting two states. Numerical methods, such as string techniques, have been employed to calculate the minimum energy paths of wetting transitions \cite{Ren2014,Pashos2015}. However, the existing theoretical and numerical studies cannot provide the complete structure of wetting transitions, primarily due to the presence of numerous local minima and saddle points.

Regarding directional liquid transport on rough surfaces, while the biomimetic experiments and development of new materials have been conducted to achieve this goal \cite{Zheng2010,Zheng2007,Chen2018,wang2021,Malvadkar2010,Chu10,Hou2023}, they have mainly concentrated on the asymmetry of contact angle hysteresis \cite{Malvadkar2010,Butt2022}. There are very limited theoretical and numerical research examining these phenomena from an energy perspective, particularly concerning the computation of multiple (meta)stable states and energy barriers involved in the movement and transfer of droplet.

In this letter, we propose a phase-field saddle dynamics method to construct the solution landscapes for wetting transitions and droplet directional transport on rough surfaces. The solution landscape serves as a pathway map that includes all stationary states of the target problem and their connections \cite{Yin2020,Yin2021}. This method provides an efficient approach to 
uncover all possible (meta)stable states and the associated transition paths for wetting transitions and directional transport. 

We first employ a phase-field function $\phi$ to characterize the liquid-vapor interface \cite{CahnHilliard}, situated within a domain $\Omega$ that features a rough boundary $\Gamma$. In the dimensionless terms, the energy $\mathcal{E}$ of the liquid-air system comprises both the bulk energy $\mathcal{E}_b$  and the surface energy $\mathcal{E}_\Gamma$, defined as follows (see, for example, \cite{XuWang2011, Qian2003}):
\begin{equation}\label{energy1}
\left\lbrace
	\begin{aligned}
     &\mathcal{E}_b(\phi) = \int_\Omega \frac{\varepsilon}{2} |\nabla \phi|^2 + \frac{1}{\varepsilon} f(\phi) d\bm{x},\\     &\mathcal{E}_\Gamma(\phi)=\int_\Gamma g(\phi) ds,
\end{aligned}
	\right.\end{equation}
where $f(\phi) = \dfrac{(\phi^2 - 1)^2}{4}$,  $g(\phi) = -\dfrac{\sigma}{4}\cos{\theta}(3\phi - \phi^3)$ with $\sigma = \dfrac{2\sqrt{2}}{3}$, $\theta$ representing Young's angle, and $\varepsilon$ being the interface thickness parameter.
As $\varepsilon$ approaches $0$, the energy $\mathcal{E}$ converges to that of the sharp interface model, which includes the energies of the liquid-air, solid-liquid, and solid-air interfaces \cite{XuWang2011}. The stable state of a droplet is determined by the following minimization problem:
\begin{equation}\label{e:minEng}
   \min \limits_{\int_\Omega \phi dx=c_0}\mathcal{E}(\phi):=\mathcal{E}_b(\phi)+\mathcal{E}_\Gamma(\phi),
\end{equation}
where $c_0$ is a constant that depends on the volume of the liquid drop. In the context of wetting on rough surfaces, there exist numerous (meta)stable states that correspond to the local minima of the energy $\mathcal{E}$.

To identify all possible stationary states, including both local minima and saddle points, we propose a phase-field saddle dynamics method to compute the index-$k$ saddle points as follows.  
\begin{equation}\label{saddle}
	\left\lbrace
	\begin{aligned}
		\partial_t\phi & = - P \delta_{\phi} \mathcal{E}(\phi) + 2 \sum\limits_{i = 1}^{k} (P\delta_\phi \mathcal{E}, u_i) u_i, \\ 
		\partial_t u_{i} & = -P \delta_\phi^2 \mathcal{E}(\phi)  u_{i} + (P \delta_\phi^2 \mathcal{E}(\phi)u_i, u_i) u_i          \\ 
		                        & \quad + 2 \sum\limits_{j=1}^{i-1} (P\delta_\phi^2 \mathcal{E}(\phi) u_{i}, u_{j})u_{j},\quad i = 1,\cdots,k,
	\end{aligned}
	\right.
\end{equation}
where $(\cdot,\cdot)$ is the $L^2$-inner product and the projection operator $P$ is defined as
\begin{equation*}
	P \phi = \phi - \frac{1}{|\Omega|} \int_\Omega \phi d\mathbf{x}.
\end{equation*}
$\delta_\phi \mathcal{E}(\phi)$ and $\delta_\phi^2 \mathcal{E}(\phi)$ are the first and second-order derivative of $\mathcal{E}(\phi)$ in the Frech\'et sense \cite{supplement}.  
The (Morse) index of a saddle point is defined by the maximal dimension of a subspace on which its Hessian is negative definite by Morse theory \cite{Morse}.
We solve this system with initial conditions  $\phi(\mathbf{x}, 0) = \phi^0(\mathbf{x})$, $\quad u_{i}(\mathbf{x}, 0) = u_{i}^0(\mathbf{x})$ such that $(u^0_{i}, u^0_{j}) = \delta_{ij}$ and  $(u_i^0, 1) = 0$ for $i,j = 1, \cdots, k.$
    
In the above system, the state variable $\phi$ moves in the ascent direction along the subspace $\mathcal{U} = {\rm \text{span}} \{u_1, u_2, \cdots, u_k\}$ and in the descent direction along its orthogonal complement. The variables $u_{i}$ $(i=1,\cdots,k)$ correspond to the first $k$ smallest eigenvalues of $P \delta_\phi^2 \mathcal{E}(\phi)$. 
It is easy to verify that the solution of \eqref{saddle} inherits the volume constraint of $\phi$ if initially  $\int_\Omega \phi_0 dx=c_0$. 
Furthermore, when $t$ becomes sufficiently large, the dynamics described by \eqref{saddle} will converge to a steady state in which the time derivatives of $\phi$ and $u_i$ approach zero. In this steady state, $\phi^*$ corresponds to the index-$k$ saddle point of $\mathcal{E}$, and $u^*_{i} (i=1,\cdots,k)$ are the $k$ unstable eigen-directions of $\phi^*$.

Using the weak form of \eqref{saddle}, we develop a finite element method for numerically computing the phase-field saddle dynamics (see supplemental materials \cite{supplement}). One significant advantage of the phase-field saddle dynamics method \eqref{saddle} is its ability to effectively manage topology changes of droplet on textured surfaces. We can then construct the solution landscape of the system via the upward/downward search algorithm — first identifying a potential highest index saddle point by using the upward search, followed by computing all connected lower index saddle points and minima via the downward search \cite{Yin2020,Yin20212,chisd}. This approach is able to provide a comprehensive structure for wetting transition and directional transport of droplet on rough surfaces as shown below.

\begin{figure*}[htbp]
\includegraphics[width=\linewidth]{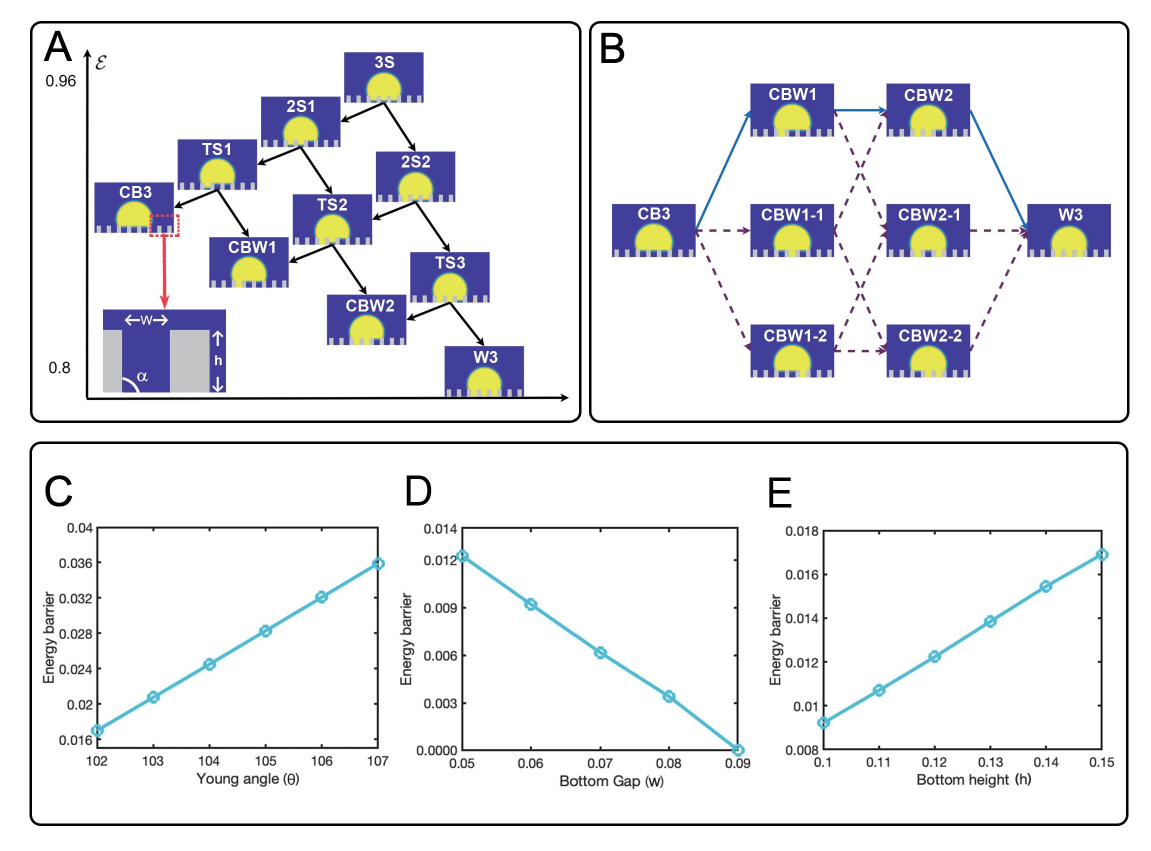}
\caption{\textbf{Wetting transition on pillared substrates}
(A) Solution landscape for the wetting transition with schematic diagrams of the micro-structures, where CB3, CBW1, CBW2 and W3 are local minima; TS1, TS2 and TS3 are index-1 saddle points; 2S1 and 2S2 are index-2 saddle points; 3S is index-3 saddle point. The inset displays the configuration of pillar substrate. (B) All transition paths from the CB3 state to the W3 state. The solid line represents the same transition path in (A) and the dash lines are equivalently alternative transition paths. (C) The energy barriers from the CB state to the Wenzel state influenced by variations in  the Young's angle $\theta$, with the gap $w=0.06$ and the base structure height $h=0.15$. (D) The energy barriers from the CB state to the Wenzel state influenced by variations in the base structure spacing $w$, with the Young's angle $\theta=102^\circ$ and the base structure height $h=0.15$.  (E) The energy barriers from the CB state to the Wenzel state influenced by variations in the bottom height $h$, with the Young's angle $\theta=102^\circ$ and the base structure gap $w=0.06$. }
\label{fig:wetting}
\end{figure*}

\textit{Wetting transition}- In the inset of Figure 1A, we display the configuration of pillar structure. The inclination angle of pillars is represented by $\alpha$, the spacing between pillars is denoted by $w$, and the height between the top and bottom of pillars is indicated by $h$. To demonstrate the functionality of the phase-field model, we examine a small droplet in two dimensions.

We first show the solution landscape for the transition of a droplet from the CB state to the Wenzel state, with the microstructure parameters of $\alpha=90^\circ$, $w=0.06$, $h=0.15$, and the Young's angle $\theta=102^\circ$ (Figure 1A). In this scenario, the index-3 saddle point, referred to as 3S, is the most unstable stationary state in the solution landscape, which can be viewed as the intermediate state between the CB state and the Wenzel state. By perturbing the 3S, we obtain two index-2 saddle points (2S1 and 2S2), three index-1 saddle points (TS1, TS2, and TS3), and four minima (CB3, CBW1, CBW2, W3). In the solution landscape, we can find a transition path from the CB state to the Wenzel state: CB3-TS1-CBW1-TS2-CBW2-TS3-W3. 
Compared to previous studies on wetting transition processes, our approach is able to determine the full transition path with all transition states and exact energy barriers. 

It is noteworthy that the symmetry of the 3S results in the perturbations applied at different positions leading to distinct resolved landscapes and transition paths from the CB3 state to the W3 state. Figure 1B illustrates all possible wetting transitions extended by Figure 1A. All these transition paths are equivalent with the same energy barrier.

With the ability to quantitatively characterize wetting transitions, we investigate how the energy barrier is influenced by the Young's angle 
$\theta$, the gap width $w$ and the height $h$ in Figure 1C-E, motivated by the results from molecular dynamics simulations \cite{Koishi09}. Figure 1C demonstrates that the energy barrier increases monotonically with an increasing Young's angle $\theta$ when $h=0.15$ and $w=0.06$. Figure 1D illustrates that the energy barrier decreases as the gap width $w$ increases with $\theta=102^\circ$ and $h=0.15$. Figure 1E shows a monotonous increase in the energy barrier with an increasing height $h$ with $\theta=102^\circ$ and $w=0.06$. 
Therefore, low hydrophobicity of the surface material, and large gap width or low altitude of the pillar structure can reduce the energy barriers and drive wetting transition, which are consistent with previous theoretical and experimental studies \cite{Bormashenko15}.

\begin{figure*}[htbp]
\includegraphics[width=\linewidth]{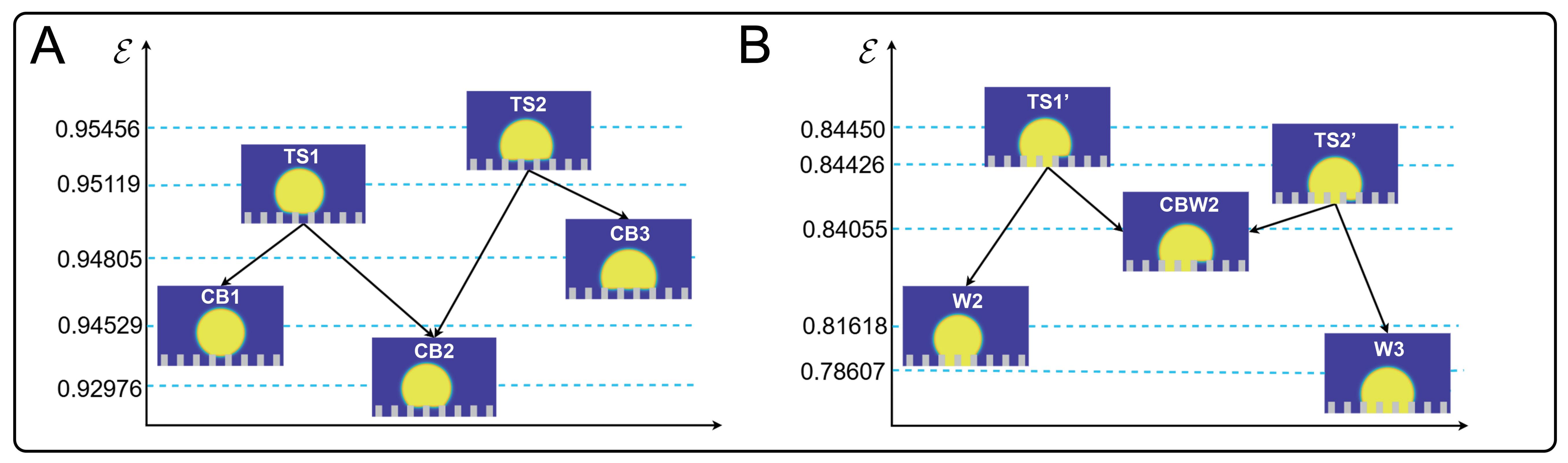}
\caption{\textbf{The transition paths between different Cassie-Baxter states and Wenzel states 
with $\theta=102^\circ$, $w=0.06$ and $h=0.15$}
(A) Transition path between different CB states where CB1,CB2 and CB3 are local minima; TS1 and TS2 are index-1 saddle points. (B) Transition path between different Wenzel states where W2, CBW2 and W3 are local minima; TS1' and TS2' are index-1 saddle points.}
\label{fig:cbcb}
\end{figure*} 

It is important to notice that there exist multiple (meta)stable states in both the liquid-filling Wenzel case and the air-trapping Cassie case. By utilizing the solution landscape, we can obtain all minima and transition states for Cassie-Baxter states and Wenzel states, respectively (Figure 2). We illustrate the transition path between three CB states—CB1, CB2, and CB3 (Figure 2A), with $\theta=102^\circ$, $w=0.06$ and $h=0.15$. Among these states, CB2 has the lowest energy, representing the desired global minimum for the Cassie-Baxter states. Similarly, we show the transition path for the Wenzel states under the same substrate configuration, denoted as W2-TS1'-CBW2-TS2'-W3, along with the global minimum W3 for the Wenzel states (Figure 2B).

\textit{Droplet directional transport}- We then utilize the phase-field saddle dynamics method to study the directional fluid motion on hydrophobic and hydrophilic rough surfaces, as observed in various experiments \cite{Zheng2007, wang2021, Malvadkar2010, Chu10}. We analyze two scenarios: the sliding of droplet on rough surfaces in Cassie-Baxter states \cite{Malvadkar2010} and in Wenzel states \cite{Chu10}. Our primary focus is on the influence of two key parameters: the Young's angle  $\theta$ and the tilt angle $\alpha$ of pillars on the substrates. For the numerical calculations that follow, we fix the gap between neighboring pillars $w=0.08$ and the height $h=0.15$.

\begin{figure*}[htbp!]
\includegraphics[width=\linewidth]{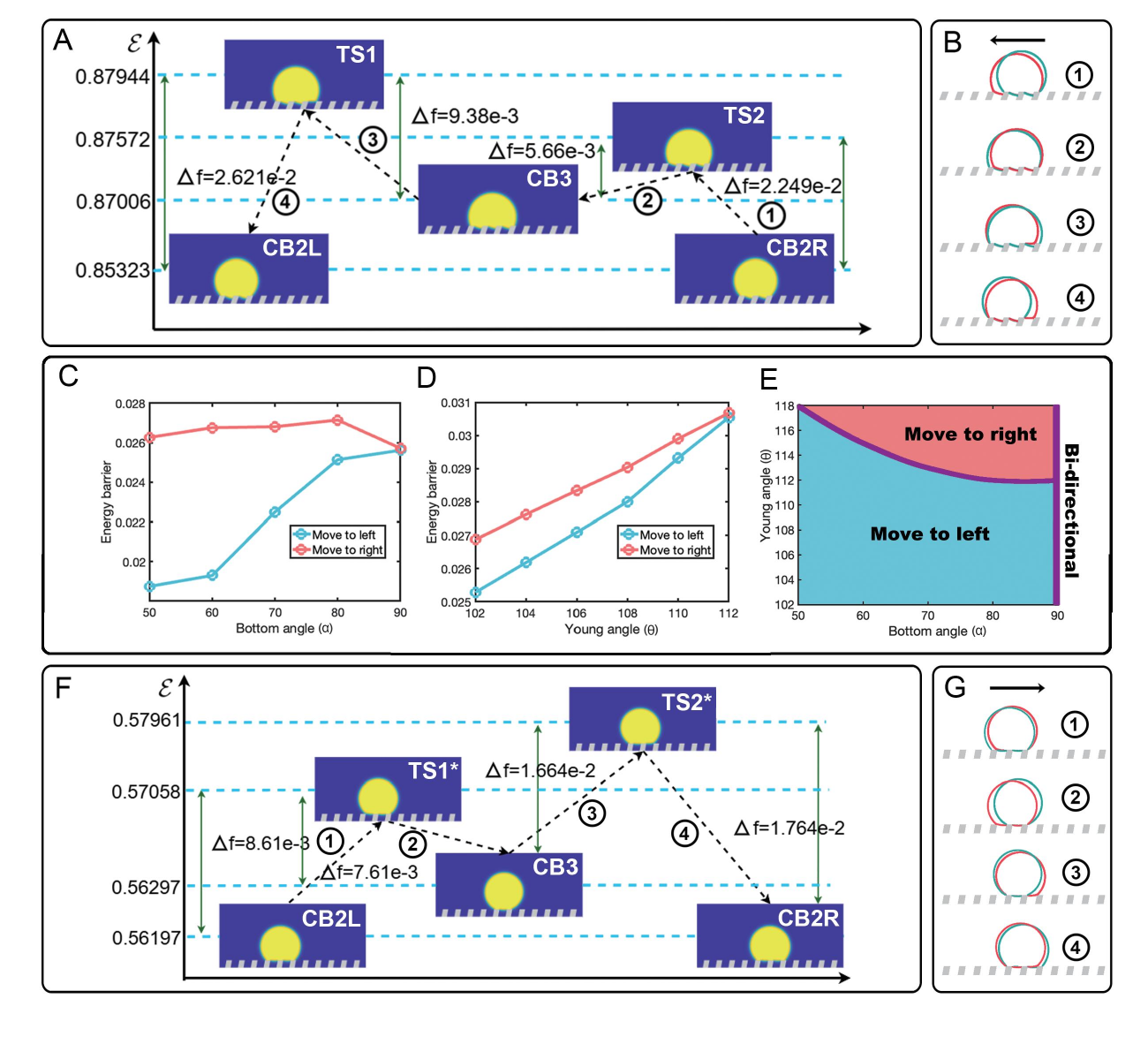}
\caption{
\textbf{Droplet directional transport on hydrophobic surfaces}
(A) Transition path for the CB state droplet transport with tilt angle $\alpha=70^\circ$, bottom surface structure spacing $w=0.08$ and surface Young's angle $\theta=102^\circ$. Wherein three (meta)stable states are CB2L, CB3 and CB2R, and two transition states are TS1 and TS2. (B) Schematic diagram illustrating the shape changes of (A), wherein the green profile represents the stable state and the red profile denotes the transition state.  (C) The energy barrier variation in different directions with respect to the tilt angles from $50^\circ$ to $90^\circ $ for fixed pillar gap $w=0.08$ and Young's angle $\theta=102^\circ$. (D) The energy barrier variation in different directions with respect to the Young's angles from $102^\circ$ to $112^\circ$ for fixed pillar gap $w=0.08$ and title angle $\theta=80^\circ$. (E) The phase diagram of droplet mobility tendencies  with respect to the Young's angle $\theta$ and the tilt angle $\alpha$. (F) The transition path for a CB state droplet transport with tilt angle $\alpha=80^\circ$, a gap $w=0.08$ and the Young's angle $\theta=114^\circ$. (G) Schematic diagram illustrating the shape changes of (F).}
\label{fig:wtrans}
\end{figure*}

We first examine droplet motion on hydrophobic surfaces in Cassie-Baxter states (Figure 3). We depict the transition path (CB2R-TS1-CB3-TS2-CB2L) for a droplet moving to the left, along with the corresponding energy barriers, with the parameters $\theta=102^\circ$ and $\alpha=70^\circ$ (Figure 3A, B). It is evident that when the droplet transitions from the left (CB2L state) to the right (CB2R state), it needs to overcome a maximum energy barrier of $\Delta f=$2.621e-2. In contrast, moving in the opposite direction only requires overcoming an energy barrier of $\Delta f=$2.249e-2. This suggests that it is easier for the droplet to move to the left rather than to the right. 

We calculate the energy barriers that a droplet needs to roll on a specific surface by varying the tilt angle with fixed $\theta=102^\circ$ (Figure 3C). We find that, when the tilt angle decreases, the energy barrier for droplet to move to left decreases and becomes lower than that for the right direction, indicating a stronger inclination to move leftward. 
However, this inclination diminishes as the Young's angle rises (Figure 3D). Notably, when the Young's angle reaches $112^\circ$, the droplet exhibits nearly bidirectional mobility on the surface structure with almost identical energy barriers for left and right movements.

This observation inspires us that we may change the direction of droplet transport by altering the Young's angle. By increasing the Young's angle $\theta=114^\circ$, the droplet's mobility tendency has changed oppositely, and the droplet's transition path becomes CB2L-TS1*-CB1-TS2*-CB2R (Figure 3F and 3G). The
energy barrier for droplet to move to right is lower than that for the left direction.
This finding is consistent with previous experimental observations that a water droplet with a contact angle of $\theta=120^\circ$ tends to move in the direction of tilting \cite{Malvadkar2010}. In Figure 3E, we provide a phase diagram that illustrates how the droplet's mobility tendencies on a rough surface depend on  Young's angle $\theta$ and tilt angle $\alpha$. The phase diagram shows the transition in the direction of droplet motion for the Cassie-Baxter states, which has not been reported in the previous literature.

\begin{figure*}[htbp]
\includegraphics[width=\linewidth]{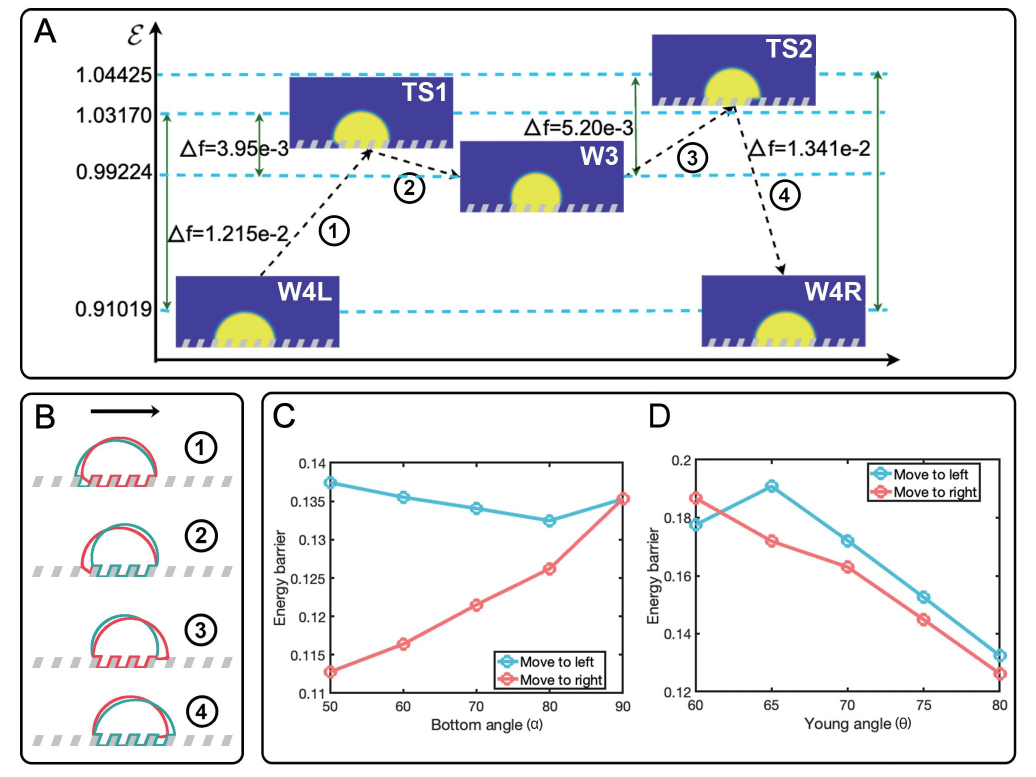}
\caption{
\textbf{Droplet directional transport on hydrophilic surfaces}
(A) Transition path for the Wenzel state droplet transport with tilt angle $\alpha=70^\circ$, bottom surface structure spacing $w=0.08$ and surface Young's angle $\theta=80^\circ$. Wherein three (meta) stable states are W4L, W3 and W4R, and two transition states are TS1 and TS2. (B) Schematic diagram illustrating the shape changes of (A), wherein the green profile represents the stable state and the red profile denotes the transition state. (C) The energy barrier variation in different directions with respect to the tilt angles from $50^\circ$ to $90^\circ$, with fixed pillar gap $w=0.08$ and Young's angle $\theta=80^\circ$. (D)The energy barrier variation in different directions with respect to the Young's angles from $60^\circ$ to $80^\circ$, with fixed pillar gap $w=0.08$ and title angle $\alpha=80^\circ$.}
\label{fig:cbtrans}
\end{figure*}

We next examine the droplet movement on hydrophilic surfaces, specifically for Wenzel states (Figure 4). We show the transition path of a droplet moving to the right, along with the corresponding energy barriers, when $\theta=80^\circ$ and $\alpha=40^\circ$ (Figure 4A). It is clear that the maximum energy barrier that the droplet moves to the right is $\Delta f=$1.215e-2, which is lower than the energy barrier for movement to the left ($\Delta f=$1.341e-2). This indicates that the droplet is more likely to move to the right on this hydrophilic surface (Figure 4B). 

We also calculate the energy barriers by changing the tilt angle $\alpha$ (Figure 4C). It is evident that 
the decrease of the tilt angle leads to the decrease of the energy barriers for moving the right, which increases the tendency for the droplet to move in the rightward direction. 
This result agrees with the intuition that the droplet is more likely to slide along the direction of tilt of the pillars, which explains the experimental findings regarding directional droplet motion on the surfaces of butterfly wings in \cite{Zheng2007}. 

We further investigate the hydrophilicity of the substrate by adjusting the Young's angle 
$\theta$ (Figure 4D). We show that the droplet tends to move to the right when the Young's angle exceeds $65^\circ$. Within this regime, as the Young's angle decreases, the energy barriers for droplet movement in both directions increase. However, when the Young's angle falls below $65^\circ$, the energy barrier for leftward movement starts to decrease and is less than the energy barrier for moving to the right at $\theta=60^\circ$. This leads to a shift in the droplet's motion tendency, which is opposite to the tilting direction. These findings offer a  theoretical explanation for the experimental observations reported in \cite{Chu10}, which describe a transition from uni-directional spreading (along the tilting direction) to bi-directional spreading within the Young's angle range of $(55^\circ,62^\circ)$ for a tilt angle of $80^\circ$.

In conclusion, we introduced a phase-field saddle dynamics method to study droplet motion on rough surfaces. This approach allows us to obtain a comprehensive solution landscape of wetting transitions and directional droplet transport on textured surfaces, encompassing all significant local minima and saddle points. 
By analyzing the landscape, we revealed the full range of Cassie-Baxter and Wenzel states, along with the complete wetting transition paths. We also quantitatively evaluated the effect of different surface design parameters on wetting transition. Furthermore, we elucidated the mechanisms of directional droplet transport on both hydrophobic and hydrophilic surfaces, demonstrating how surface design can influence these behaviors.

From the theoretical and computational perspective, the proposed approach provides an efficient computational tool and a unifying view of wetting transition and directional transport. Our methodology can be readily generalized to three-dimensional scenarios. Additionally, the phase-field model can easily incorporate the effects of vapor dissolution in the liquid, allowing us to quantify the wetting transition under high liquid pressure \cite{Lv2014,Xiang2017}. 

\section*{ACKNOWLEDGMENTS}
L.Z. was supported by the National Natural Science Foundation of China (No.12225102, T2321001, and 12288101). X.X. was supported by National Natural Science Foundation of China (No.11971469 and 12371415) and Beijing Natural Science Foundation (No.Z240001). Y.Z. was supported by the National Natural Science Foundation of China (No.12301554),  the Natural Science Foundation of Jiangsu Province (No.BK20220368), Natural science fund for colleges and universities in Jiangsu Province (No.22KJD110003).


\section*{DECLARATION OF INTERESTS}
The authors declare no competing interests.

\section*{LEAD CONTACT WEBSITE}
\url{http://faculty.bicmr.pku.edu.cn/~zhanglei/}

\bibliographystyle{apsrev4-2}

\end{document}